\documentclass[preprint,12pt]{elsarticle}

\usepackage{amsmath, amssymb}
\usepackage{natbib}
\usepackage[labelfont=bf]{caption}
\usepackage{subcaption}
\usepackage{graphicx}
\usepackage{hyperref}
\usepackage{booktabs}

\usepackage{tikz}
\usetikzlibrary{patterns}
\tikzset{
  mycircle/.style={draw, circle, minimum size=1cm},
  mysquare/.style={draw, rectangle, minimum size=1cm},
  myellipse/.style={draw, ellipse, minimum width=2cm, minimum height=1cm},
  myarrow/.style={->, thick},
  myroundrect/.style={draw, rounded corners, minimum width=2cm, minimum height=1cm}
}

\journal{Computers \& Geoscie}

\begin{document}

\begin{frontmatter}

\title{Neural Networks for Predicting Permeability Tensors of 2D Porous Media: Comparison of Convolution- and Transformer-based Architectures}

\author[inst1]{Sigurd Sønvisen Vargdal}
\author[inst1]{Paula Reis}
\author[inst2]{Henrik Andersen Sveinsson}
\author[inst1,ify]{Gaute Linga}


\affiliation[inst1]{organization={PoreLab, The Njord Centre, Department of Physics, University of Oslo},
            city={Oslo},
            country={Norway}}

\affiliation[ify]{organization={PoreLab, Department of Physics, Norwegian University of Science and Technology},
            city={Trondheim},
            country={Norway}}

\affiliation[inst2]{organization={The Njord Centre, Department of Physics},
            institution={University of Oslo},
            city={Oslo},
            country={Norway}
            }

\begin{abstract}
    Permeability is a central concept in the macroscopic description of flow through porous media, with applications spanning from oil recovery to hydrology. 
    Traditional methods for determining the permeability tensor involving flow simulations or experiments can be time consuming and resource-intensive, while analytical methods, e.g., based on the Kozeny-Carman equation, may be too simplistic for accurate prediction based on pore-scale features.
    In this work, we explore deep learning as a more efficient alternative for predicting the permeability tensor based on two-dimensional binary images of porous media, segmented into solid ($1$) and void ($0$) regions.
    We generate a dataset of 24,000 synthetic random periodic porous media samples with specified porosity and characteristic length scale. 
    Using Lattice-Boltzmann simulations, we compute the permeability tensor for flow through these samples with values spanning three orders of magnitude.
    We evaluate three families of image-based deep learning models: ResNet (ResNet-$50$ and ResNet-$101$), Vision Transformers (ViT-T$16$ and ViT-S$16$) and ConvNeXt (Tiny and Small).
    To improve model generalisation, we employ techniques such as weight decay, learning rate scheduling, and data augmentation. 
    The effect of data augmentation and dataset size on model performance is studied, and we find that they generally increase the accuracy of permeability predictions.
    We also show that ConvNeXt and ResNet converge faster than ViT and degrade in performance if trained for too long.
    ConvNeXt-Small achieved the highest $R^2$ score of $0.99460$ on $4,000$ unseen test samples. 
    These findings underscore the potential to use image-based neural networks to predict permeability tensors accurately.
\end{abstract}


\end{frontmatter}

\section{Introduction}
Predicting the transport properties of porous media is crucial to many industries, such as oil recovery, groundwater management, and CO$_2$ subsurface storage \citep{Feder_Flekkøy_Hansen_2022,DENTZ20111}.
Slow, steady flow through porous media is described macroscopically by Darcy's law: 
\begin{equation}\label{eq:darcy}
    \mathbf{u} = - \frac{\mathsf{K}}{\mu } \left(\nabla p - \mathbf{f}\right),
\end{equation}
where $\mathbf{u}$ is the average fluid velocity, $\mu$ is the fluid viscosity, $p$ is the pressure, and $\mathbf{f}$ is a body force such as that due to gravity.
The permeability tensor $\mathsf{K}$ is a purely geometric quantity that encodes small-scale structure and relates pressure gradients to flow rates. In effect, it quantifies how easily a fluid flows through a porous medium.
Low permeability is often associated with complex, tortuous structures, while high permeability indicates more open and direct flow paths. 

Determining the permeability accurately is not straightforward. Traditionally, experiments or simulations of the flow through the pore-scale microstructure are needed, and these can be time-consuming and resource-intensive. Theoretical estimates, such as those based on the Kozeny--Carman equation, tend to be less accurate on complex structures because they do not incorporate key geometric information 
\citep{Feder_Flekkøy_Hansen_2022,carman1956}. 
Pore-network models \citep{BENNOAH2024104809} provide a computationally cheaper alternative, but their accuracy depends strongly on how faithfully the pore geometry and connectivity are represented; simplified network constructions often lead to reduced predictive accuracy.
The inaccuracy of theoretical estimates and the fact that experimental and numerical measurements are demanding and must be done on a case-by-case basis, motivate the search for alternative, efficient and accurate methods.

The success of convolutional neural networks (CNNs) in image recognition \citep{726791}, scaled with AlexNet in 2012 \citep{NIPS2012_c399862d} and extended by ResNet in 2015 \citep{resnet}, demonstrated the potential for data-driven feature extraction. 
Later, the transformer architecture \citep{NIPS2017_3f5ee243} enabled the Vision Transformer (ViT) \citep{vit}, which models long-range dependencies using self-attention and tokenization of images instead of convolutions. 
More recently, ConvNeXt \citep{convnext} has bridged CNNs and transformers by incorporating architectural refinements such as patchification, large kernels, and inverted bottlenecks, along with modern training techniques. These changes enable convolution-based models to achieve performance comparable to transformers.

In recent years, image-based deep learning has emerged as a promising alternative for predicting permeability. 
The accuracy of the prediction is often measured with the coefficient of determination:
\begin{equation*}
    R^2 = 1 - \frac{\sum_{n=0}^{N-1} ||\mathsf{\hat{K}}_n - \mathsf{K}_n||_F^2}{\sum_{n=0}^{N-1} ||\mathsf{\hat{K}}_n - \mathsf{\bar{K}}||_F^2},
\end{equation*}
where we denote the Frobenius norm of a $d \times d$ matrix $\mathsf{A}$ as $|| \mathsf{A} ||_F = \sum_{i=1}^d \sum_{j=1}^d A_{ij}$. $\mathsf{K}$ is the predicted permeability tensor, $\mathsf{\hat{K}}$ is the target permeability tensor, and $\mathsf{\bar{K}}$ is the mean of the target permeability values.
\citet{Araya-Polo2020} combined high-resolution two-dimensional (2D) images obtained from thin sections of core plugs with laboratory measurements of their permeabilities. They achieved $R^2=0.9582$ on thin-section subsamples, and $R^2=0.7967$ when evaluating across larger spans of the core plug scans.
\citet{Graczyk_2020} applied CNNs to predict porosity, permeability, and tortuosity from synthetic 2D Poisson porous media, where the flow-dependent properties were determined for a single flow direction using Lattice-Boltzmann simulations. 
\citet{TAKBIRI2022110069} adopted a U-Net \citep{u-net} architecture to predict velocity fields in simple geometric domains (ellipses and rectangles) and derived permeability from them. 
On a test set with discs as obstacles their model achieved the highest $R^2=0.98$ on the $K_{xx}$ component.
\citet{zhaiImprovedConvolutionalNeural2024} compared ConvNeXt, ViT, DensNet \citep{densnet}, and ResNet in a single flow direction for permeability prediction. They used 2D synthetic images and a finite-element solver to obtain the flow fields. On a test set of 200 images, their ConvNeXt achieved the highest accuracy with $R^2=0.9667$.
\citet{kashefi2025visionmambapermeabilityprediction} applied the Vision Mamba \citep{zhu2024visionmambaefficientvisual} to single direction permeability prediction in 3D, achieving $R^2=0.9969$ on $170$ synthetic test samples. Their total dataset consisted of $1,692$ samples with a porosity range of $[0.125, 0.200]$, and their flow fields were based on Lattice-Boltzmann simulations.
\citet{Avilkin2025} evaluated an XGBoost \citep{Chen2016XGBoost} model using geometric features from 2D slices of synthetic 3D porous structures, using Lattice-Boltzmann simulations to calculate the permeability. Depending on the slice configuration, they reported $R^2$ scores of $0.79$ (single slice), $0.85$ (three random slices), and up to $0.91$ when using three orthogonal slices.

In this work, we evaluate the capability of three architectures---ResNet, ViT, and ConvNeXt---to predict the full permeability tensor of complex 2D porous media samples.
As modern deep learning methods typically benefit from larger and more diverse datasets \citep{scalinglawslanguage}, we here increase both the dataset size and the model capacity compared to previous studies, aiming for higher predictive accuracy.
In particular, we generate $24,000$ synthetic random porous media samples ($128$ by $128$ pixels) spanning three orders of magnitude in (principal) permeability, which we determine using Lattice-Boltzmann simulations.
The $24,000$ samples are split into three groups; $16,000$ for training, $4,000$ for validation and model selection, and $4,000$ for testing as the final benchmark.
To further increase the effective dataset size, we develop data augmentation techniques based on the symmetries of the problem.
To improve generalisation, we employ regularization techniques including weight decay, learning rate scheduling, and gradient clipping.
We find that our highest test score $R^2=0.99460$ is achieved using ConvNeXt-Small, and that the ViT architecture requires more training and more data to achieve comparable results to the convolution-based architectures.

The paper is structured as follows: First, we go through how the synthetic samples are configured and simulated in section \ref{sec:data}. Secondly, we show the configuration of the models and what numerical tests we conduct to find the models best performance in section \ref{sec:nn}. Thirdly, we show our results and discuss them in section \ref{sec:r&d}. Lastly, we give our conclusions in section \ref{sec:conc}.

\section{Data generation}\label{sec:data}
\subsection{Porous media geometry}
We generate synthetic porous media samples that are generic, easy to generate, smooth, and that contain structures like grains and dead-end pores as seen in thin sections of natural porous media.
This is achieved by generating a $128 \times 128$ array of pseudo-random numbers between 0 and 1 (using the generator PCG64 \citep{prng}), which is then smoothed using a 2D Gaussian filter with characteristic structure length $\sigma = 4.0$. Binary solid-fluid fields are obtained by thresholding the continuous field values ($\in [0, 1]$) at a sought porosity $\phi$, with the range $\phi \in [0.001, 0.9]$.

After generation, the medium may contain disconnected fluid clusters---fluid regions that are completely enclosed by solids and therefore isolated from the main flow. Examples of such clusters can be seen along the vertical edges of the domain in Figure~\ref{fig:example_media}. To reduce computational cost, these media are post-processed by filling in the disconnected fluid regions, yielding what we refer to as \textit{filled images}. Because flow simulations are restricted to the fluid domain, removing the isolated clusters reduces the number of fluid nodes and therefore the computational load. Moreover, since disconnected regions contain no flow, they do not contribute to the average velocity in Equation~\ref{eq:darcy} and hence do not affect the permeability. For the Kozeny–Carman calculations, the filled images are used, as they represent the effective porosity of the connected medium.
To implement this method, we modify Scikit-Image's \citep{scikit} \texttt{binary\_blobs} function to enforce periodic boundary conditions.

\subsection{Flow configuration}
Our samples are generated with single-phase laminar flow in a two-dimensional domain, driven by a body force representing gravity. No-slip boundary conditions are applied between fluid and solid regions, and periodic boundary conditions are used along the domain walls. 

We solve for Stokes flow, given by
\begin{align*}
    &\mu \nabla^2 \mathbf{u}_{ss} - \nabla p_{ss} + \mathbf{f} = \mathbf{0},\\
    & \nabla \cdot \mathbf{u}_{ss} = 0,
\end{align*}
where $\mathbf{u}_{ss}$ is the small scale fluid velocity, ${p}_{ss}$ is the small scale pressure, and $\mathbf{f}$ is a body force.
To compute the permeability tensor, we use the global average 
$\mathbf{u}$ of the small scale fluid velocity $\mathbf{u}_{ss}$:
\begin{align}
    \mathbf{u} = \frac{1}{A_f} \int_{\Omega_f} \mathbf{u}_{ss} \, d^2 \mathbf{x},
\end{align}
where $A_f$ is the area of the fluid domain $\Omega_f$.
The same is for $p$ and $p_{ss}$. 

We compute the permeability using Darcy's law (Equation~\ref{eq:darcy}). Since the domain is periodic, the pressure gradient averages to zero. On matrix form the permeability is given by:
\begin{equation*}
    \mathsf{K} = \mu f^{-1}\begin{bmatrix}
        \mathbf{u}^{(x)}&\mathbf{u}^{(y)}
    \end{bmatrix},
\end{equation*}
where $\mathbf{u}^{(x)}$ is the resulting average fluid velocity vector from a body force in the in the $x$ direction, i.e., $\mathbf{f} = f \hat{\mathbf x}$ and $\mathbf{u}^{(y)}$ is the global average fluid velocity resulting from $\mathbf{f} = f \hat{\mathbf y}$. A more explicit simplification of equation \ref{eq:darcy} is shown in appendix \ref{app:darcy}.

\subsection{Lattice-Boltzmann simulations}
Flow simulations are performed using the Lattice Boltzmann Method (LBM) with a single-relaxation-time (BGK) collision  \citep{Feder_Flekkøy_Hansen_2022}. The evolution of the distribution function $N_i(\mathbf{x}, t)$ is given by
\begin{equation*}
    N_i(\mathbf{x} + \mathbf{c}_i \Delta t, t + \Delta t) 
    = N_i(\mathbf{x}, t) 
    - \omega \left[N_i(\mathbf{x}, t) - N_i^{eq}(\mathbf{x}, t)\right] + F_i,
\end{equation*}
where $\omega$ is the relaxation rate, $\mathbf{c}_i$ are the discrete lattice velocities, and $N_i^{eq}$ is the local equilibrium distribution, $F_i$ is the body force. 

A $D2Q9$ lattice is employed, which uses nine discrete velocity directions (including the rest particle $\mathbf{c}_0 = [0,0]$). This lattice provides a good balance between numerical stability and computational cost for two-dimensional incompressible flow.


All quantities are expressed in lattice units. This yields permeability values of order unity, which is convenient for training neural networks. Each simulation is run for 10,000 time steps to ensure convergence to steady-state flow. 

\begin{figure}
    \centering
    \includegraphics{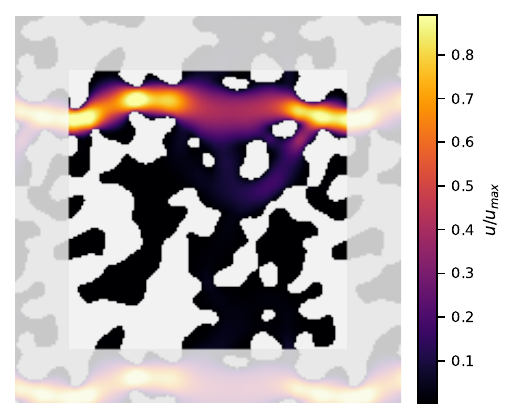}
    \caption{A synthetic porous medium with periodic geometry. The colouring indicates relative flow speed. The shaded area contains parts of periodic images of the domain.}
    \label{fig:example_media}
\end{figure}
\begin{figure*}
    \centering
    \includegraphics{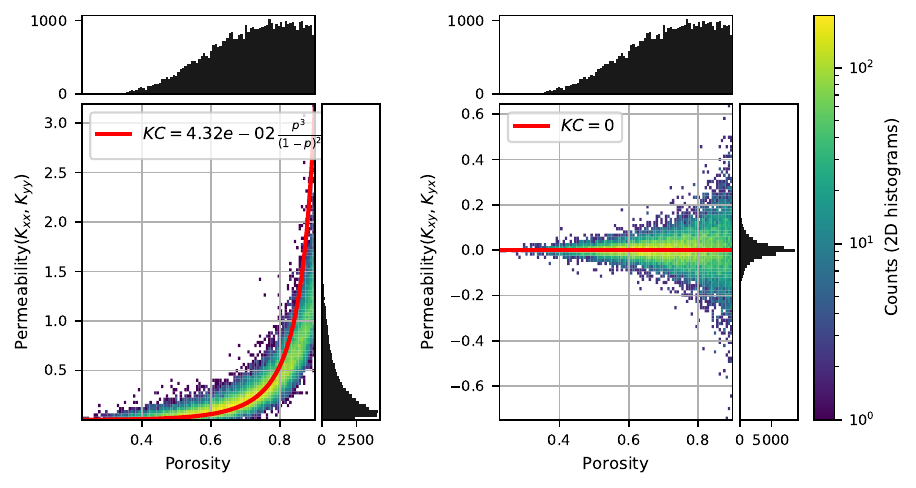}
    \caption{Distribution of permeability component values over porosity for all $24,000$ samples. The permeability tensor values are $K_{ij}$ for $i,j=x$ or $y$. $KC$ is the Kozeny-Carman fit to the data.}
    \label{fig:perm_vs_porosity}
\end{figure*}

Figure~\ref{fig:example_media} shows an example of a synthetic porous medium generated with this procedure. Figure~\ref{fig:perm_vs_porosity} shows the distribution of permeability tensor components $K_{ij}$ over porosity for all $24,000$ samples. The values of principal permeability (largest eigenvalue of the tensor) span three orders of magnitude.

\section{Neural networks}\label{sec:nn}
\subsection{Models}
We seek to train a neural network model to predict the permeability tensor from a 2D image. We explore three main architectures with different size variants. 
For ResNet, we train two versions, ResNet-$50$ and ResNet-$101$, shown in Table \ref{tab:resnet_architectures}. The ResNet-Stem block is given circular padding as our system is periodic. 
The different configurations we train for ViT and ConvNeXt are shown, respectively in Table \ref{tab:vit_architectures} and \ref{tab:convnext_architectures}.

We do not use dropout \citep{dropout} or stochastic depth \citep{stocasticdepth}.
All tests are implemented in PyTorch \citep{pytorch} and models are trained from a cold start using the default weight initialization, which uses a uniform distribution bounded by $1/\sqrt{\text{input features}}$, on NVIDIA RTX A4000, Tesla P100, and A100 GPUs with half-precision (FP16) training. 

Our models are trained with mean square error (MSE) loss:
\begin{equation*}
    \mathrm{MSE} = \frac{1}{N} \sum_{i=0}^{N-1} ||\mathsf{\hat{K}}_i - \mathsf{K}_i ||_F^2,  
\end{equation*}
where $N$ is the number of samples per batch, $\mathsf{\hat{K}}$ is the target permeability values, and $\mathsf{K}$ is the predicted permeability values.

\begin{table}
    \centering
    \caption{ResNet architectures. Depth refers to the number of residual blocks in each stage. Width is the number of channels in each stage. The final width equals the input width to the fully connected layer.}
    \begin{tabular}{lcc}
        \toprule
        \textrm{Model} & \textrm{Depth} & \textrm{Width} \\
        \midrule
        ResNet-$50$  & $\{3,4,6,3\}$  & $\{64,256,512,1024,2048\}$ \\
        ResNet-$101$ & $\{3,4,23,3\}$ & $\{64,256,512,1024,2048\}$ \\
        \bottomrule
    \end{tabular}
    \label{tab:resnet_architectures}
\end{table}

\begin{table}
    \centering
    \caption{ViT architectures. Patch size is the pixel width/height each image is divided into. The embedding dim is the width of the embedding layer. Depth is the number of transformer blocks. Heads is the number of attention heads in each block. MLP ratio is the width of the inverse bottleneck compared to the input.}
    \begin{tabular}{lccccc}
    \toprule
    \textrm{Model} & \textrm{Patch Size} & \textrm{Embed Dim} & \textrm{Depth} & \textrm{Heads} & \textrm{MLP Ratio} \\
    \midrule
    ViT-S$16$ & $16$ & $384$ & $12$ & $6$  & $4$ \\
    ViT-T$16$ & $16$ & $192$ & $12$ & $3$  & $4$ \\
    \bottomrule
    \end{tabular}
    \label{tab:vit_architectures}
\end{table}

\begin{table}
    \centering
    \caption{ConvNeXt architectures. Width is the number of channels in each stage. Depth is the number of ConvNeXt blocks in each stage. The patch size in the stem layer, and the MLP ratio was 4 for both models.}
    \begin{tabular}{lcc}
        \toprule
        \textrm{Model} & \textrm{Width} & \textrm{Depth} \\
        \midrule
        ConvNeXt-Tiny  & $\{96,192,384,768\}$ & $\{3,3,9,3\}$ \\
        ConvNeXt-Small & $\{96,192,384,768\}$ & $\{3,3,27,3\}$ \\
        \bottomrule
    \end{tabular}
    \label{tab:convnext_architectures}
\end{table}

\subsection{Numerical studies}
We conduct a series of numerical studies to assess the models' generalization behaviour and performance under different training conditions.
Firstly, we study how data augmentation affects the models’ ability to generalize from training to validation samples. 
In the context of porous media, augmentation corresponds to applying geometric symmetries such as rotations and reflections. 
Each transformation produces a new valid realization of the same physical system, where the permeability tensor $\mathsf{K}$ transforms accordingly as
\begin{equation}\label{eq:permut}
    \mathsf{K}' = \mathsf{P} \mathsf{K} \mathsf{P}^T,
\end{equation}
where $\mathsf{P}$ is a two by two permutation matrix representing the applied transformation.
To systematically study the effect of each augmentation type, we build up the transformations incrementally:  
starting with only horizontal flips (H), then only vertical flips (V), then rotations (R) by $90^\circ$, $180^\circ$, and $270^\circ$. 
Next, we combine flips (H+V), and finally include both flips and rotations (H+V+R).  
Both H and V are sampled independently with probability $50\%$. All rotations are sampled with $1/3$ probability for each.
The complete set of unique rotations and reflections of a square forms the dihedral group $D_4$, which contains eight symmetry operations. 
By sampling these transformations uniformly, each original image can generate up to eight unique variants, effectively multiplying the number of distinct training samples by eight. 
H+V+R can reach all eight unique variants, H and V add two new variants, and R adds 3 new variants.
\begin{figure}
    \centering
    \begin{tikzpicture}[node distance=1.8cm and 2.2cm, every node/.style={inner sep=0}]
        \node (id) at (0,0) {\includegraphics[scale=0.2]{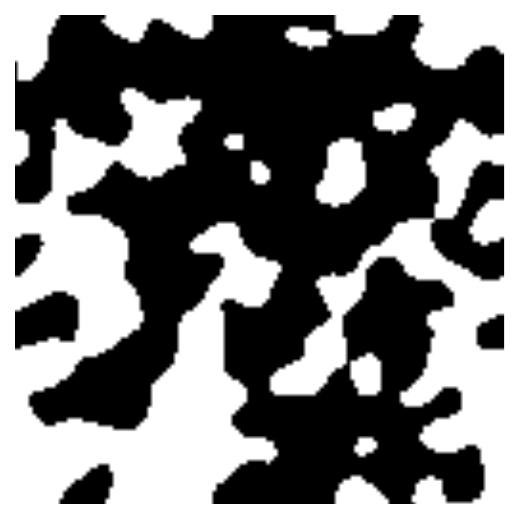}};
        \node at (0,-1) {Id};

        \node (r90) [rotate=90] at (2,0) {\includegraphics[scale=0.2]{demo_media.pdf}};
        \node at (2,-1) {$R_{90}$};

        \node (r180) [rotate=180] at (4,0) {\includegraphics[scale=0.2]{demo_media.pdf}};
        \node at (4,-1) {$R_{180}$};
        
        \node (r270) [rotate=270] at (6,0) {\includegraphics[scale=0.2]{demo_media.pdf}};
        \node at (6,-1) {$HR_{270}$};
        
        \node (h) [yscale=-1] at (0,-2.5) {\includegraphics[scale=0.2]{demo_media.pdf}};
        \node at (0,-3.5) {$H$};

        \node (hr) [yscale=-1, rotate=90] at (2,-2.5) {\includegraphics[scale=0.2]{demo_media.pdf}};
        \node at (2,-3.5) {$HR_{90}$};

        \node (hr2) [yscale=-1, rotate=180] at (4,-2.5) {\includegraphics[scale=0.2]{demo_media.pdf}};
        \node at (4,-3.5) {$HR_{180}$};

        \node (hr3) [yscale=-1, rotate=270] at (6,-2.5) {\includegraphics[scale=0.2]{demo_media.pdf}};
        \node at (6,-3.5) {$HR_{270}$};
    \end{tikzpicture}
    \caption{The 8 elements of the dihedral group $D_4$ acting on the synthetic porous medium. 
    }
    \label{fig:D4_images}
\end{figure}
Figure~\ref{fig:D4_images} shows the eight symmetry operations of $D_4$ applied to an example image.
The augmentation configurations used in this study are summarized in Table~\ref{tab:aug_schemes}.
\begin{table}
    \centering
    \caption{Augmentation schemes and their definitions.}
    \begin{tabular}{ll}
        \toprule
        \textrm{Augmentation} & \textrm{Description} \\
        \midrule
        None            & No augmentation \\
        Horizontal Flip & $50\%$ chance horizontal reflection \\
        Vertical Flip   & $50\%$ chance vertical reflection \\
        Rotation        & $R_{90}, R_{180}, R_{270}$ (each with probability $33.3\%$) \\
        H+V Flip        & Horizontal ($50\%$) and vertical ($50\%$) flip \\
        H+V+Rotate      & H+V flips and $R_{90}, R_{180}, R_{270}$ \\
        $D_4$           & All elements with probability $1/8$ \\
        \bottomrule
    \end{tabular}
    \label{tab:aug_schemes}
\end{table}
The explicit permeability permutations are shown in appendix \ref{app:perm}.

Secondly, we investigate how model performance depends on the number of training samples.
Models are trained on datasets ranging from 2,000 to 20,000 samples in increments of 2,000, with an $80\%$–$20\%$ split between training and validation samples. The 4,000 test samples are held out for final model comparison in Section~\ref{sec:bench}.
We perform this test on both ConvNeXt and ViT, using the Tiny and Small configurations.

Third, we investigate the effect of training duration. Since the models differ in size and architecture, they may require different numbers of epochs to converge. Training for too many epochs can lead to overfitting and reduced performance. To study this, we train all models described above across a range of epochs and analyse their convergence behaviour.

Finally, we present a full benchmark that compares the best-performing configurations of each architecture: ConvNeXt-Tiny and ConvNeXt-Small, ViT-S$16$ and ViT-T$16$, and ResNet-$50$ and ResNet-$101$. 
For reference, we also include predictions from the Kozeny–-Carman equation, which provides a classical permeability–porosity estimate:
\begin{equation*}
    K = C\frac{\phi^3}{(1-\phi)^2},
\end{equation*}
where  $\phi$ is the porosity and $C$ is the mean value when using the training data to calculate $C=K_{ii}(1-\phi)^2/\phi^2$.

Additionally, we confirm the validity of $R^2$ as a metric of accuracy through similarity plots of the Kozeny-Carman equation, the neural network model with the lowest $R^2$, and the neural network model with the highest $R^2$.

Unless stated otherwise, all models were trained with the hyperparameters listed in Table~\ref{tab:hyperparams_for_training}. 
\begin{table}
    \centering
    \caption{Hyperparameters used for training. Some values vary depending on receptive test.}
    \begin{tabular}{lc}
        \toprule
        \textrm{Hyperparameter} & \textrm{Value} \\
        \midrule
        Optimizer & AdamW\citep{AdamW} \\
        Epochs & 500 \\
        Base Learning rate & 0.0008 \\
        Learning rate decay & Cosine \\
        Weight decay & 0.1 \\
        Augmentation & $D_4$ \\
        Warmup & Linear \\
        Warmup steps & 1000 \\
        Training samples & 16000 \\
        Validation samples & 4000 \\
        Test samples & 4000 \\
        Batch size & 128 \\
        \bottomrule
    \end{tabular}
    \label{tab:hyperparams_for_training}
\end{table}

\section{Results and discussion}\label{sec:r&d}
\subsection{Effect of data augmentation}
\begin{table}
    \centering
    \caption{Comparison of ConvNeXt-Small and ViT-S$16$ under different data augmentations. Validation set $R^2$ and MSE are shown.}
    \begin{tabular}{lcc|cc}
        \toprule
        \textrm{Augmentation} & \multicolumn{2}{c}{\textrm{ConvNeXt-Small}} & \multicolumn{2}{c}{\textrm{ViT-S$16$}} \\
        & \textrm{$R^2$} & \textrm{MSE} & \textrm{$R^2$} & \textrm{MSE} \\
        \midrule
        None            & 0.99086 & 0.000818 & 0.93991 & 0.005387 \\
        Horizontal Flip & 0.99490 & 0.000465 & 0.95502 & 0.004067 \\
        Vertical Flip   & 0.99473 & 0.000475 & 0.95487 & 0.004069 \\
        Rotate        & 0.99750 & 0.000229 & 0.96715 & 0.002974 \\
        H+V Flip        & 0.99735 & 0.000241 & 0.97132 & 0.002614 \\
        H+V+Rotate      & 0.99885 & 0.000107 & 0.99862 & 0.000134 \\
        $D_4$           & 0.99898 & 0.000096 & 0.99859 & 0.000138 \\
        \bottomrule
    \end{tabular}
    \label{tab:aug_results}
\end{table}
Table \ref{tab:aug_results} presents the validation $R^2$ scores and MSE values for the different combinations of data augmentation applied to ConvNeXt-Small and ViT-S$16$. Increasing the amount of augmentation leads to higher $R^2$ scores and lower MSE values. Combining multiple augmentations further improves performance. ViT-S$16$ shows the largest gain, with a difference of $\Delta R^2 = 0.05898$ between the unaugmented and $D_4$ configurations. ConvNeXt-Small starts from a higher baseline $R^2$ without augmentation and achieves a slightly better final $R^2$ with $D_4$ than ViT-S$16$.

The most likely explanation for these gains is that the augmentations add new artificial but physically valid samples to the dataset. Horizontal and vertical flips each introduce one additional variation per sample, effectively doubling the number of training samples from $16,000$ to $32,000$. Although horizontal and vertical flips only add one new variation when applied separately, they add three unique variations when combined.
Rotations add three variation per sample, while $H+V+R$ and $D_4$ adds eight variations of each sample. 
Unlike in classification tasks, the parallel to flipping and rotating is less direct here. When an augmentation is applied in this regression setting, the corresponding label is also transformed, meaning that each augmented instance represents a new valid sample rather than a slightly perturbed copy of the original.

There is a clear benefit to combining augmentations. For ConvNeXt-Small, the gain from combining flips is similar to that achieved with rotations, while ViT-S$16$ shows a slightly larger improvement. Our results further show that the $D_4$ configuration provides a similar performance gain to $H+V+R$, likely because both generate the same eight variations when sampled repeatedly. The small difference between them is most likely due to training noise.

We observe that ViT-S$16$ has the highest increase in performance from data-augmentation. The reason for this could be that the transformer architecture requires more samples to achieve high accuracy, compared to convolution-based architectures. The ViT has a weaker inductive bias and tends to need more data and augmentation/regularization to achieve high accuracy. As discussed by \citet{convnext}, much of the ViT’s performance arises from improved training techniques and scaling of both model and dataset size. Our findings support this interpretation.

\subsection{Effect of dataset size}
\begin{figure}
    \centering
    \includegraphics{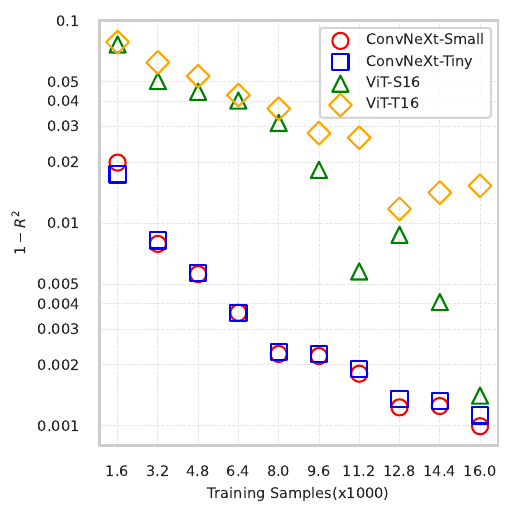}
    \caption{Plot of validation $1-R^2$ score for ConvNeXt and ViT in the tiny and small configuration over different sizes of the dataset.}
    \label{fig:diff_num_datapoints}
\end{figure}
Figure \ref{fig:diff_num_datapoints} shows the validation $1-R^2$ score for ConvNeXt-Small/Tiny and ViT-S/T$16$ when trained on different numbers of training samples. We see a clear trend that models improve when given more data to train on, as suggested by the added variations of data augmentation. The ConvNeXt models reach a high accuracy even for smaller dataset sizes and is close to saturated after $12,800$ when trained on samples. The ViT models require more data than ConvNeXt to achieve comparable accuracy, which only ViT-S$16$ reaches on all $16,000$ training samples. 

\subsection{Convergence study}\label{sec:convergence}
\begin{figure}
    \centering
    \includegraphics{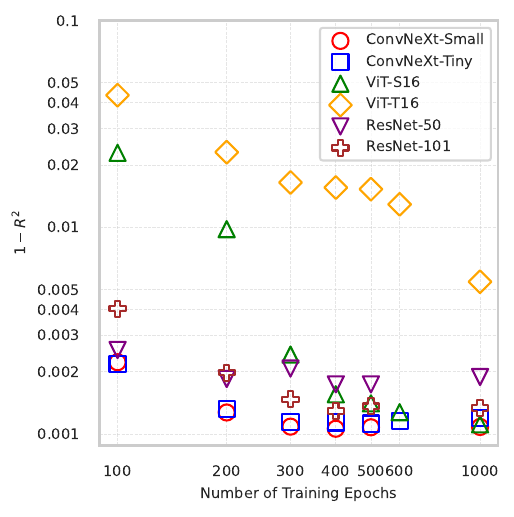}
    \caption{Best (lowest) $1-R^2$ value obtained for each training run over the total number of  epochs the model was trained for.}
    \label{fig:diff_num_epochs}
\end{figure}
Figure \ref{fig:diff_num_epochs} shows the best (lowest) $1-R^2$ value obtained for each training run, plotted against the total number of epochs for which the model was trained. We observe that models tend to be more accurate when trained for longer. For some models, training too long decreases performance. ResNet-$50$ and ConvNeXt-Tiny perform worse when trained for more than $500$ epochs. ResNet-$101$ and ConvNeXt-Small get worse after $400$ epochs. 
ViT-S$16$ achieves similar performance to ConvNeXt-Small when trained for $1,000$ epochs. ViT-T$16$ sees a large relative increase in accuracy when trained to $1,000$ epochs. \citet{scalinglawslanguage} showed that smaller transformers need to process more tokens than larger transformers to achieve comparable performance. ViT-T$16$'s jump from $600$ epochs to $1,000$ epochs suggests that, had it been trained for longer than $1,000$, it might have achieved comparable performance to ViT-S$16$. Additional investigations of convergence during training are shown in appendix \ref{app:converge}.

\subsection{Best achieved performance across models}\label{sec:bench}
\begin{table}
    \centering
    \begin{tabular}{lcccc}
        \toprule
        \textrm{Model} & \textrm{Val $R^2$} & \textrm{Test $R^2$} & \textrm{Epoch Run} & \textrm{Epoch}\\
        \midrule
        ConvNeXt-Small & \textbf{0.99894} & \textbf{0.99460} & 400 & 369 \\
        ViT-S$16$      & 0.99889 & 0.99308 & 1000 & 995 \\
        ConvNeXt-Tiny  & 0.99887 & 0.99433 & 500 & 498 \\
        ResNet-$101$   & 0.99871 & 0.99287 & 400 & 380 \\
        ResNet-$50$    & 0.99826 & 0.99078 & 500 & 436 \\
        ViT-T$16$      & 0.99455 & 0.96672 & 1000 & 990 \\
        Kozeny-Carman      & - & -0.22402 & - & - \\
        \bottomrule
    \end{tabular}
    \caption{Best achieved validation $R^2$ and corresponding test $R^2$, for each models best performing training length. The best values are shown in bold.
    }
    \label{tab:best_acc}
\end{table}
Table \ref{tab:best_acc} shows how the models performed with the  validation and test $R^2$ score.
ConvNeXt-Small achieves $R^2=0.99894$ on the validation set after 369 epochs when trained to 400 epochs, which is the highest among the tested models. On the test set it achieves $R^2=0.99460$, which is the highest out of our models. 
ConvNeXt-Tiny performs similarly to ConvNeXt-Small and while ViT-S$16$ and the ResNets have similar performance on the validation set, the $R^2$ drops significantly on the test set. 
ViT-T$16$ has the largest drop in $R^2$ from validation to test data. A possible explanation for this could be that the model has not fully converged yet, as discussed in Section~\ref{sec:convergence}.

To confirm that $R^2$ is a reliable accuracy metric, we show similarity plots of one of the best and one of the worst performing models in terms of $R^2$. We also show the respective similarity based on the Kozeny-Carman fit.
\begin{figure*}
    \centering
    \includegraphics[width=0.99\textwidth]{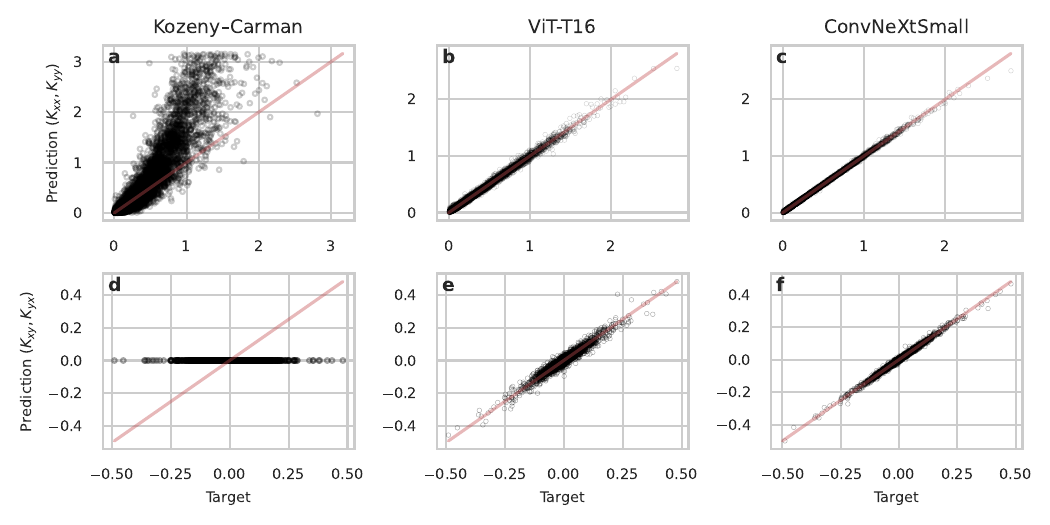}
    \caption{Similarity plot of Kozeny-Carman, ViT-T$16$, and ConvNeXt-Small. Panels \textbf{a-c} shows the similarity for the diagonal components $K_{xx}$ and $K_{yy}$; panels \textbf{d-f} show the similarity for the off diagonal components $K_{xy}$ and $K_{yx}$.
    The identity line where Target = Prediction is shown in red.}
    \label{fig:sim}
\end{figure*}
Figure~\ref{fig:sim} shows the similarity between the predicted permeability and the respective target permeability on the test set, where Figure~\ref{fig:sim} a \& d shows the similarity for the Kozeny-Carman fit.
Kozeny-Carman gives no value for cross-directional permeability components $K_{yx}$ and $K_{xy}$, and so it predicts these terms to be zero. 
The geometric parameter $C$ that we obtain from the fit is $C_{xx}= 4.33\cdot 10^{-2}$ and $C_{yy}= 4.35\cdot 10^{-2}$, while $C_{xy}$ and $C_{yx}$ are defined to be zero.
For one of the lowest $R^2$ models, Figure~\ref{fig:sim} b \& e shows ViT-T$16$. We observe that the model has learnt to predict the overall trend of all the components of permeability.
For one of the highest $R^2$ models, Figure~\ref{fig:sim} c \& f shows ConvNeXt-Small. The model shows a strong capability to predict all components with tiny outliers for larger permeability values. 
With this observation, we confirm that $R^2$ is a reliable metric of accuracy in this regression problem.

We observe that some of the models appear to have different exact numerical performances in some tests. This variability between training runs can likely be attributed to stochastic effects (non-deterministic GPU kernels) and numerical differences in floating-point arithmetic across hardware platforms. For example, training on an Nvidia P100 executes mixed-precision in fp32 fallback, while an A100 uses native fp16 Tensor Core operations. 

\section{Conclusion}\label{sec:conc}
In this work, we investigated the use of image-based deep learning models to predict the permeability tensor of 2D periodic porous media. By generating 24,000 synthetic porous medium samples and using Lattice-Boltzmann simulations to calculate their permeability, we trained and evaluated three deep learning architectures: ResNet, ViT, and ConvNeXt. 
$4,000$ samples were withheld from the training as final test samples.

Various data-augmentation techniques, including horizontal and vertical flips, rotations, and their combination based on $D_4$ were tested. We found that increasing the number of variants introduced by augmentation generally improves the accuracy of the models. The $D_4$ and the combination of horizontal flips, vertical flips, and rotations achieved similar results up to noise in the training, likely due to new variations added by the augmentations.

Our results show that model performance increases with more data. For ConvNeXt the performance is close to saturated after $10,000$ samples. ViT-S$16$ achieved similar performance to ConvNeXt when trained on the full dataset. ViT-T$16$ shows signs of insufficient training for more than $16,000$ samples.

We observed that different models converge differently. Models with fast convergence can perform poorer if trained too long. ResNets and ConvNeXts achieved their best performance if trained to $400$-$500$ epochs. ViTs converge beyond $1000$ epochs.
We find that while the ViT architecture can predict permeability with high accuracy, convolution-based architectures are easier to train and require less data to achieve good performance. ConvNeXt-Small is our most accurate model.

Our results suggest that deep learning provides a fast and accurate alternative to traditional simulations for estimating permeability tensors from images.
Future work should explore how the models perform on larger domains, such as larger grids or anisotropic 3D media where the permeability tensor will have significant off-diagonal components.

\section{Acknowledgements}
This work has received support from the Research Council of Norway through their FRINATEK funding scheme, projects 325819 (M4) and 354100 (ChemFrac) and through their Center of
Excellence funding scheme, project no.\ 262644 (PoreLab).

\section{Code Availability}
The code to generate data, train the models, and produce the plots are available at: \href{https://github.com/Tensorboy2/permeability-prediction/tree/main}{https://github.com/Tensorboy2/permeability-prediction/tree/main}. 

\section{Data Availability}
The dataset and the best preforming model weights are available at \href{https://zenodo.org/records/17711512}{https://zenodo.org/records/17711512}.

\appendix

\section{Permeability from Darcy's law}\label{app:darcy}
Darcy's law in matrix form is given by equation \ref{eq:darcy}. With the periodic boundaries used in our flow simulations, the pressure gradient term in the equation vanishes. This gives us:
\begin{equation*}
    \begin{bmatrix}
        \mathbf{u}^{(1)} & \mathbf{u}^{(2)}
    \end{bmatrix} = \frac{\mathsf{K}}{\mu} \begin{bmatrix}
        \mathbf{f}^{(1)} & \mathbf{f}^{(2)}
    \end{bmatrix},
\end{equation*}
where $(1)$ denotes the simulation with force in the x direction and $(2)$ denotes simulation with force in the y direction.
We can simplify the force term as the forcing strength times the identity matrix. Then we can rearrange the terms as:
\begin{equation*}
    \mathsf{K}=\mu f^{-1}
    \begin{bmatrix}
        \mathbf{u}^{(1)} & \mathbf{u}^{(2)}
    \end{bmatrix}.
\end{equation*}
In component form we then get:
\begin{align*}
    & K_{xx} = \mu f^{-1}u_{x}^{(1)}\\
    & K_{xy} = \mu f^{-1}u_{y}^{(1)}\\
    & K_{yx} = \mu f^{-1}u_{x}^{(2)}\\
    & K_{yy} = \mu f^{-1}u_{y}^{(2)}\\
\end{align*}

\section{Augmentations}\label{app:perm}
The general expression for the permutation of the permeability tensor $K$ by a permutation matrix $P$ is given by Eq \ref{eq:permut}.
We use in total 5 different permutations, defined as:
\begin{align*}
    &90^o\text{ rotation}:R_{90}= 
    \begin{pmatrix}
        0 & -1\\
        1 & 0
    \end{pmatrix},\\
    &180^o\text{ rotation}:R_{180}= 
    \begin{pmatrix}
        -1 & 0\\
        0 & -1
    \end{pmatrix},\\
    &270^o\text{ rotation}:R_{270}= 
    \begin{pmatrix}
        0 & 1\\
        -1 & 0
    \end{pmatrix},\\
    &\text{Horizontal flip}:H=\begin{pmatrix}
        -1 & 0\\
        0 & 1
    \end{pmatrix},\\
    &\text{Vertical flip}:V=\begin{pmatrix}
        1 & 0\\
        0 & -1
    \end{pmatrix}.
\end{align*}
Applying these permutations yield the following:
\begin{align*}
    &\text{Horizontal flip: } 
    \begin{pmatrix}
        k_{xx} & k_{xy}\\
        k_{yx} & k_{yy}
    \end{pmatrix}
    \rightarrow 
    \begin{pmatrix}
        k_{xx} & -k_{xy}\\
        -k_{yx} & k_{yy}
    \end{pmatrix}\\
    &\text{Vertical flip: } 
    \begin{pmatrix}
        k_{xx} & k_{xy}\\
        k_{yx} & k_{yy}
    \end{pmatrix}
    \rightarrow 
    \begin{pmatrix}
        k_{xx} & -k_{xy}\\
        -k_{yx} & k_{yy}
    \end{pmatrix}\\
    &\text{Rotation by 90$^\circ$: }
    \begin{pmatrix}
        k_{xx} & k_{xy}\\
        k_{yx} & k_{yy}
    \end{pmatrix}
    \rightarrow 
    \begin{pmatrix}
        k_{yy} & -k_{yx}\\
        -k_{xy} & k_{xx}
    \end{pmatrix}\\
    &\text{Rotation by 180$^\circ$: }
    \begin{pmatrix}
        k_{xx} & k_{xy}\\
        k_{yx} & k_{yy}
    \end{pmatrix}
    \rightarrow 
    \begin{pmatrix}
        k_{xx} & k_{xy}\\
        k_{yx} & k_{yy}
    \end{pmatrix}\\
    &\text{Rotation by 270$^\circ$: }
    \begin{pmatrix}
        k_{xx} & k_{xy}\\
        k_{yx} & k_{yy}
    \end{pmatrix}
    \rightarrow 
    \begin{pmatrix}
        k_{yy} & -k_{yx}\\
        -k_{xy} & k_{xx}
    \end{pmatrix}
\end{align*}

Eight unique variations can be achieved with combinations of $H$ and $R_{90}$. Written out this becomes:
\begin{align*}
    &I = 
    \begin{pmatrix}
        1 & 0\\
        0 & 1
    \end{pmatrix},\\
    &R_{90} =
    \begin{pmatrix}
        0 & -1\\
        1 & 0
    \end{pmatrix},\\
    &R_{180} =
    \begin{pmatrix}
        -1 & 0\\
        0 & -1
    \end{pmatrix},\\
    &R_{270} =
    \begin{pmatrix}
        0 & 1\\
        -1 & 0
    \end{pmatrix},\\
    &H =
    \begin{pmatrix}
        -1 & 0\\
        0 & 1
    \end{pmatrix},\\
    &HR_{90} =
    \begin{pmatrix}
        0 & 1\\
        1 & 0
    \end{pmatrix},\\
    &HR_{180} =
    \begin{pmatrix}
        1 & 0\\
        0 & -1
    \end{pmatrix},\\
    &HR_{270} =
    \begin{pmatrix}
        0 & -1\\
        -1 & 0
    \end{pmatrix}.
\end{align*}

For a general $K$ this becomes:
\begin{align*}
    &\text{Identity } I: &
    \begin{pmatrix} k_{xx} & k_{xy} \\ k_{yx} & k_{yy} \end{pmatrix} 
    &\rightarrow
    \begin{pmatrix} k_{xx} & k_{xy} \\ k_{yx} & k_{yy} \end{pmatrix},\\[2mm]
    &\text{Rotation } R_{90}: &
    \begin{pmatrix} k_{xx} & k_{xy} \\ k_{yx} & k_{yy} \end{pmatrix} 
    &\rightarrow
    \begin{pmatrix} k_{yy} & -k_{yx} \\ -k_{xy} & k_{xx} \end{pmatrix},\\[2mm]
    &\text{Rotation } R_{180}: &
    \begin{pmatrix} k_{xx} & k_{xy} \\ k_{yx} & k_{yy} \end{pmatrix} 
    &\rightarrow
    \begin{pmatrix} k_{xx} & k_{xy} \\ k_{yx} & k_{yy} \end{pmatrix},\\[2mm]
    &\text{Rotation } R_{270}: &
    \begin{pmatrix} k_{xx} & k_{xy} \\ k_{yx} & k_{yy} \end{pmatrix} 
    &\rightarrow
    \begin{pmatrix} k_{yy} & -k_{yx} \\ -k_{xy} & k_{xx} \end{pmatrix},\\[2mm]
    &\text{Horizontal flip } H: &
    \begin{pmatrix} k_{xx} & k_{xy} \\ k_{yx} & k_{yy} \end{pmatrix} 
    &\rightarrow
    \begin{pmatrix} k_{xx} & -k_{xy} \\ -k_{yx} & k_{yy} \end{pmatrix},\\[2mm]
    & H R_{90}: &
    \begin{pmatrix} k_{xx} & k_{xy} \\ k_{yx} & k_{yy} \end{pmatrix} 
    &\rightarrow
    \begin{pmatrix} k_{yy} & k_{yx} \\ k_{xy} & k_{xx} \end{pmatrix},\\[2mm]
    & H R_{180}: &
    \begin{pmatrix} k_{xx} & k_{xy} \\ k_{yx} & k_{yy} \end{pmatrix} 
    &\rightarrow
    \begin{pmatrix} k_{xx} & -k_{xy} \\ -k_{yx} & k_{yy} \end{pmatrix},\\[2mm]
    & H R_{270}: &
    \begin{pmatrix} k_{xx} & k_{xy} \\ k_{yx} & k_{yy} \end{pmatrix} 
    &\rightarrow
    \begin{pmatrix} k_{yy} & k_{yx} \\ k_{xy} & k_{xx} \end{pmatrix}.
\end{align*}

\section{Reduced patch size}
The Vision Transformer (ViT) architecture was originally developed for 224×224-pixel images, where a patch size of 16 corresponds to 196 tokens per image. In our experiments, the input resolution is 128×128, which results in only 64 tokens for the same patch size and thus a lower computational load compared to ConvNeXt of similar scale. To investigate whether the reduced number of tokens limits performance, we test smaller patch sizes to increase token count and capture finer-grained spatial information.
Specifically, we evaluate the configurations ViT-T8 and ViT-S8, which use a patch size of 8 and produce 256 tokens per image. For ViT-S8, the batch size was reduced to 64 to fit within GPU memory. We acknowledge that gradient accumulation could have bypassed this.

\begin{figure}
    \centering
    \includegraphics{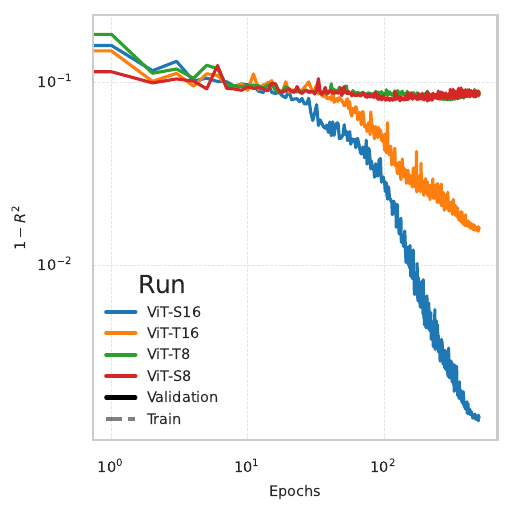}
    \caption{Validation $1-R^2$ over epochs in log-log scale for ViT tiny and small with patch sizes $16$ and $8$. Optimal result for $1-R^2$ is to go to 0.}
    \label{fig:reduced patch size}
\end{figure}
Figure \ref{fig:reduced patch size} shows the validation $1-R^2$ over epochs for ViT-T/S with patch sizes 8 and 16. Reducing the patch size from 16 to 8 led to a complete loss of learning in both models. This indicates that the improvement trends observed in prior classification studies do not generalize to our permeability prediction task.

The lack of learning with smaller patches can be explained by the limited information contained in our binary single-channel images, compared to the RGB natural images used by \citet{wang2025scalinglawspatchificationimage}. In our case, reducing patch size likely results in patches containing too little local information for the model to learn meaningful representations. Similar findings are discussed in \citet{kashefi2025visionmambapermeabilityprediction}, who provide a more comprehensive analysis of patch size effects in permeability prediction using Vision Mamba.

\section{Additional convergence study}\label{app:converge}
\begin{figure}
    \centering
    \includegraphics[scale=0.65]{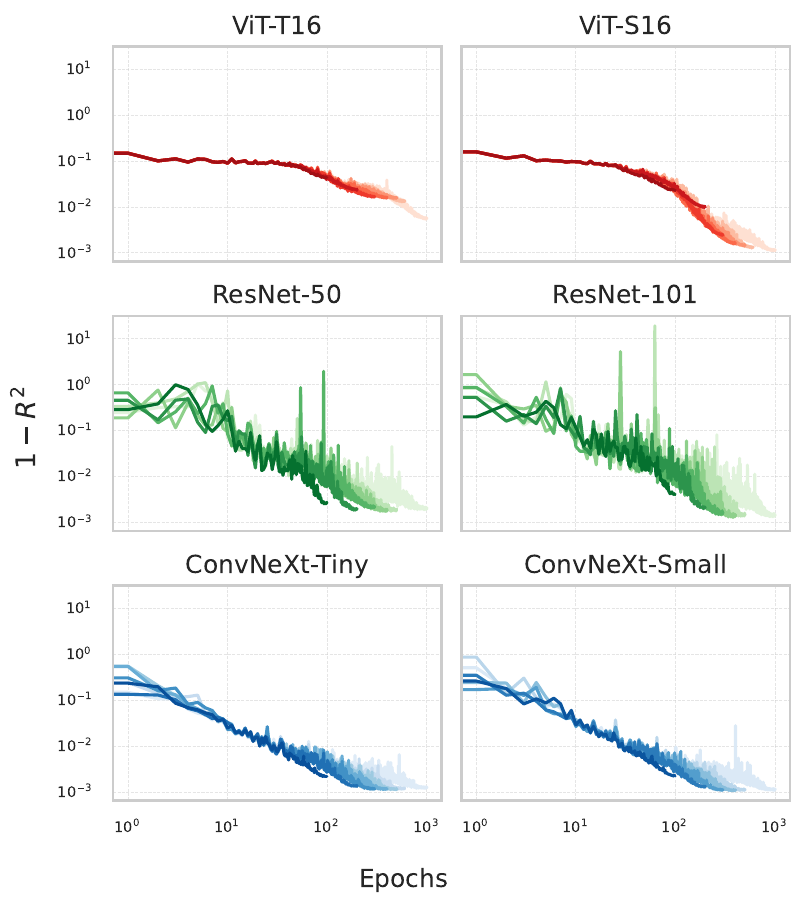}
    \caption{Validation $1-R^2$ over epochs for training models up to $1000$ epochs.}
    \label{fig:all_models_row}
\end{figure}
Figure \ref{fig:all_models_row} shows the validation $1-R^2$, illustrating the training of each model for different ranges of total epochs. It shows that convolution-based models converge faster than ViT architectures on this regression problem. 
We observe that ConvNeXts has a much more stable training than ResNets, which has much more jumps in validation $1-R^2$ values. The possible reason for this could be the training technique. Although we used the same training techniques for all models, the ResNets are normally not trained with either warmup or a decay scheduler on the learning rate. \citet{resnet} divided the learning rate by $10$ for 2 specific epochs and did not use further tuning. The final results of the models are satisfactory, so we do not explore the learning rates of ResNets further.

\section{Transformer scaling}
\begin{figure}
    \centering
    \includegraphics{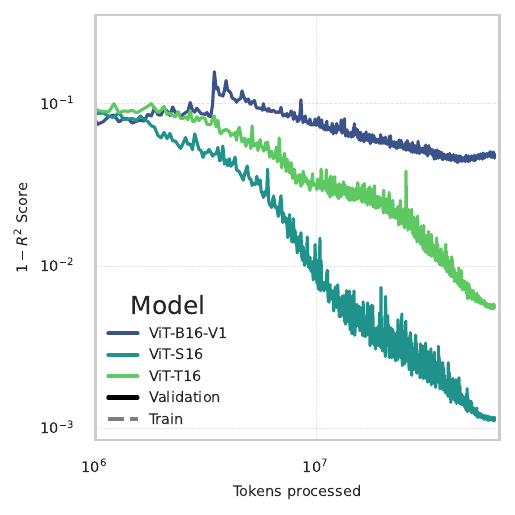}
    \caption{$1-R^2$ for training ViT-T/S/B$16$ up to $1000$ epochs over tokens processed.}
    \label{fig:longer_vits}
\end{figure}
Our results suggest that ViTs require much longer training as they only see improved accuracy when trained longer. Figure \ref{fig:longer_vits} shows $1-R^2$ for ViT-$16$ in 3 different sizes Tiny, Small, and Base when trained to $1000$ epochs. 

When \citet{scalinglawslanguage} studied the scaling of natural language processing (NLP) with the transformer, they found that larger models, while demanding more computation per batch, require fewer tokens processed to be more accurate than their smaller variants. Our results show a similar behaviour for the Small and Tiny model, but the Base model does not fit this criterion. This suggests diminishing returns from scaling ViTs for this regression problem.

\bibliographystyle{elsarticle-num-names}
\bibliography{references}
\end{document}